\documentstyle[12pt]{article}

\begin{document}
\title{\textbf{Slow plasma dynamo driven by electric current helicity in non-compact Riemann surfaces of negative curvature}} \maketitle
{\textbf{L.C. Garcia de Andrade}-Departamento de F\'{\i}sica
Te\'orica-IF-UERJ-RJ, Brasil\\[-0.1mm]
\vspace{0.1cm} \paragraph*{Boozer addressed the role of magnetic helicity in dynamos [Phys Fluids \textbf{B},(1993)]. He pointed out that the magnetic helicity conservation implies that the dynamo action is more easily attainable if the electric potential varies over the surface of the dynamo. This provided us with motivation to investigate dynamos in Riemannian curved surfaces [Phys Plasmas \textbf{14}, (2007);\textbf{15} (2008)]. Thiffeault and Boozer [Phys Plasmas (2003)] discussed the onset of dissipation in kinematic dynamos. When curvature is constant and negative, a simple simple laminar dynamo solution is obtained on the flow topology of a Poincare disk, whose Gauss curvature is $K=-1$. By considering a laminar  plasma dynamo [Wang et al, Phys Plasmas (2002)] the electric current helicity ${\lambda}\approx{2.34m^{-1}}$ for a Reynolds magnetic number of $Rm\approx{210}$ and a growth rate of magnetic field $|{\gamma}|\approx{0.022}$. Negative constant curvature non-compact $\textbf{H}^{2}$, has also been used in one-component electron 2D plasma by Fantoni and Tellez (Stat Phys, (2008)). Chicone et al (CMP (1997)) showed fast dynamos can be supported in compact $\textbf{H}^{2}$. PACS: 47.65.Md. Key-word: dynamo plasma.}
\section{Introduction}
Earlier Boozer \cite{1} has investigated magnetic helicity driven dynamos, where the magnetic helicity constraint is enhanced if the electric potential varies over surface of the dynamo. He argues that in the case of the Earth, the north-south pole variation posseses an electric potential which varies a hundred volts. Recently helicity constraints have also been investigated by Thiffeault and Boozer \cite{2} where the dissipation is taken into account. In their case, they found that helicity generation terms are exponential smaller than energy dissipation, so that large amounts of energy are dissipated before the any helicity can be created. In this paper, use as made of the Riemannian geometry of Cauchy metric in the chaotic plasma flows, where the magnetic field is stretched in the plasma flow \cite{3}. In their case the high conducting fluid, with high magnetic Reynolds numbers Rm of the order $10^{8}-10^{15}$, and consequently very low dissipation. \newline
Here one addresses the converse issue and considers the case of a non-ideal plasma where the dynamo action survives on the Riemannian manifold of negative constant curvature in the form of a Lobachevsky plane. The geodesic flows in the plasma are computed using the geodesic equation in 3D. This geometry in the form of a paraboloid can be easily shown to focusing the magnetic field orthogonal to its surface. It is important to stress that this does not happens in the Euclidean plane, or the spherical surface where the magnetic field lines orthogonal to their respective surfaces remain parallel or diverge. This provides also another strong fountain of motivation for investigating the magnetic flows in geodesic dynamos in Riemannian spaces of negative constant curvature. Since as shown by Chicone et al \cite{4} even fast dynamos can be supported in Riemannian compact 2D manifolds of constant negative curvature, the Cowling anti-fast dynamo theorem for 2D surfaces is not violated here. \newline
Anti-fast dynamo theorems have also been addressed by Garcia de Andrade \cite{5}. The slow dynamo flows obtained here are shear flows, which can also be obtained by the stretch-fold-shear dynamo mechanism investigated previously by Bayly, Childress \cite{6} and Gilbert \cite{7}. In this paper the absence of advection terms is due to the presence of a comoving term which makes the spatial flow vanishes. This kind of frame is very well known in cosmology and can be used in near future to investigate cosmological dynamos. The dynamo flow used here is certainly more complex that the simple uniform stretching dynamo flow investigated by Arnold et al \cite{8}. Another motivation for the use of negative constant Riemann curvature dynamo plasma surfaces, has been the one-component two-dimensional plasma by Fantoni and Tellez \cite{9} in the realm of electron plasmas in 2D. Here as happens in general relativity the plasma undergoes a Coriolis force which is given by the presence of the curvilinear coordinates effects present in the Riemann-Christoffel symbol in the MHD dynamo equation. In their non-relativistic plasma limit, Fantoni and Tellez \cite{9}. have used a Flamm's paraboloid, which is a non-compact manifold which represents the spatial Schwarzschild black hole, to investigate one-component two-dimensional plasmas. \newline
Restoring forces and magnetic field reversals possibility are also discussed in 3D slow dynamo curved surfaces. Recently another sort of slow dynamos in liquid sodium laboratory has been modelling by Shukurov at al \cite{10}, by embedding a Moebius strip flow in the three-dimensional space. The paper is organized as follows: Section II presents the mathematical formalism necessary to grasp the rest of the paper. In the next section the slow dynamo solution is presented as well as the non-geodesic equations is computed. In this section III, the sign of magnetic helicity in the exponential growth of the slow dynamo is shown to be important to the slow dynamo action. Both helicities are computed  on the hyperbolic Poincare disk. Discussions and conclusions are presented in section IV.
\newpage
\section{Slow dynamo plasmas in curved surfaces}
In the Euclidean three-dimensional space $\textbf{E}^{3}$ describe by Lobachevsky plane geometry can be presented here for the benifit of non-mathematical inclined reader. The Lobachevsky metric is given by
\begin{equation}
ds^{2}=y^{-2}[dx^{2}+dy^{2}]
\label{1}
\end{equation}
where ${\textbf{H}}^{2}=(w=x+iy;y>0)$ is the hyperbolic plane in its half-upper part. Here $\sqrt{-1}=i$ is the imaginary unit of the complex plane $\textbf{C}$. The Ricci tensor and Riemann-Christoffel symbols of the Lobachevsky metric
\begin{equation}
R_{11}=\frac{1}{y^{2}}\label{2}
\end{equation}
\begin{equation}
R_{22}=\frac{1}{y^{2}}\label{3}
\end{equation}
\begin{equation}
R=2\label{4}
\end{equation}
\begin{equation}
{{\Gamma}^{1}}_{21}={{\Gamma}^{2}}_{22}=-\frac{1}{y}\label{5}
\end{equation}
\begin{equation}
{{\Gamma}^{2}}_{11}=\frac{1}{y}\label{6}
\end{equation}
Riemann curvature tensor is given by
\begin{equation}
{R_{1212}}=-\frac{1}{y^{4}}\label{7}
\end{equation}
The Kretschmann scalar invariant, so much used in GR to determine whether a singularity is not a true singularity or an horizon, just in Schwarzschild black hole geometry, is given by
\begin{equation}
{\cal{R}}={R_{1212}}{R^{1212}}=-1\label{8}
\end{equation}
which shows that the line $y=0$ represents a fake singularity or an event horizon of the 2D section of the universe. The process by which the particles are stretched in the plasma flow to give rise to dynamo, is the geodesic equation
\begin{equation}
\frac{d^{2}J}{ds^{2}}+K(s)J=0
\label{9}
\end{equation}
whose solution for the negative curvature hyperbolic space is
\begin{equation}
J(s)=J_{0}sinh(\sqrt{-K}s)=J_{0}sinh(s)
\label{10}
\end{equation}
Note that the force-free dynamo equation yields
\begin{equation}
{\Delta}B=-curl(curl B)=-{\lambda}^{2}B
\label{11}
\end{equation}
where
\begin{equation}
curl B={\lambda}B
\label{12}
\end{equation}
is the force-free Beltrami equation. From the assumption that the comoving frame is using one obtains
\begin{equation}
curl[v{\times}B]=0
\label{13}
\end{equation}
The expression for the self-induction equation
\begin{equation}
\dot{\textbf{B}}={\eta}{\Delta}\textbf{B}+{\nabla}{\times}[\textbf{v}{\times}\textbf{B}]
\label{14}
\end{equation}
where ${\Delta}={\nabla}^{2}$ is the Laplacian in general curvilinear coordinates. Therefore the calculation of this term shall be fundamental in our case. Let us expand this term in terms of Cartesian coordinates Laplacian
\begin{equation}
{\Delta}_{Flat}={{\partial}_{x}}^{2}+{{\partial}_{y}}^{2}\label{15}
\end{equation}
and the Riemann-Christoffel connection
\begin{equation}
{{\Gamma}^{i}}_{jk}=\frac{1}{2}g^{il}[g_{lj,k}+g_{lk,j}-g_{jk,l}]\label{16}
\end{equation}
where $(i,j=1,2,3)$. Now let us consider the above MHD dynamo equation in curvilinear coordinates. Since the advection term is in principle not present, our first worry should be to compute the first
\begin{equation}
{\Delta}\textbf{B}=\frac{1}{\sqrt{g}}{\partial}_{i}(\sqrt{g}g^{ij}{{\partial}_{j}}\textbf{B})
\label{17}
\end{equation}
Here, ${\gamma}$, the rate of the amplification of the magnetic field from the ansatz
\begin{equation}
\textbf{B}=\textbf{B}_{0}(\textbf{x})e^{{\gamma}t}\label{18}
\end{equation}
The covariant expression for the Laplacian operator then becomes
\begin{equation}
{\Delta}=[{\partial}_{i}g^{ij}+{\Gamma}^{j}]{\partial}_{j}+{{\nabla}_{F}}^{2}
\label{19}
\end{equation}
here ${\nabla}_{F}$ is the flat gradient in Cartesian $(x,y)$ coordinates. Here
\begin{equation}
{\Gamma}^{i}:=g^{ij}{\Gamma}_{j}=Tr({{\Gamma}^{i}}_{jk})\label{20}
\end{equation}
is the trace of the above Riemann-Christoffel symbol. To derive the expression (\ref{19}) one used the Riemannian geometry identity for the trace of Riemann-Christoffel system
\begin{equation}
{\Gamma}_{i}:=\frac{1}{\sqrt{g}}{\partial}_{i}[\sqrt{g}]\label{21}
\end{equation}
By taking the solenoidal constraint on the magnetic field $divB=0$ one obtains the form of the field as
\begin{equation}
B^{i}={B^{i}}_{0}e^{{\gamma}t}y^{2}\label{22}
\end{equation}
Note that this expression shows that unless the y coordinate is bounded the magnetic field grows spatially without bounds. Since the only constraint on y is that it be positive, this certainly may be the case. If one uses the Riemann-Christoffel connections of the above Lobachevsky-Poincaré hyperbolic disk, one may find that the first to terms in the general Laplace-Beltrami operator By using the force-free condition above one obtains the dynamo equation as
\begin{equation}
{\gamma}B^{i}=-{\eta}{\lambda}^{2}B^{i}
\label{23}
\end{equation}
The Maxwell magnetic two-form F is
\begin{equation}
F:=F_{ij}dx^{i}{\wedge}dx^{j}=B_{x}dy{\wedge}dz+B_{y}dz{\wedge}dx+B_{z}dx{\wedge}dy
\label{24}
\end{equation}
where ${\wedge}$ symbol means the wedge skew-symmetric product. Note that the $B^{z}$ is the component of the magnetic field orthogonal to the Lobachevsky plane. Focusing of the negative curvature surface magnetic field can be done by the orthogonal magnetic fields to its surface. A simple drawing of the paraboloid can show that these magnetic lines converge to some focusing point outside the paraboloid. It is easy to check that the expression (\ref{21}) yields a solution for the dynamo equation (\ref{22}) as long as
\begin{equation}
{\gamma}=-{\eta}{\lambda}^{2}
\label{25}
\end{equation}
which shows that by the slow dynamo condition
\begin{equation}
lim_{{\eta}\rightarrow{0}}\textbf{Re}{\gamma}(\eta)=0
\label{26}
\end{equation}
Of course in fast dynamos expression (\ref{26}) would be positive. Here $\textbf{Re}$ represents the real part of the growth rate scalar ${\gamma}$. Note that from this expression the constraint ${\lambda}^{2}\ge{0}$ implies that either the dynamo slowliness is enhanced or the dynamo is marginal $({\gamma}=0)$. This result is obtained since the slow dynamo criteria predominates over magnetic field decay.
\section{Electric and magnetic helicities and force-free slow dynamos}
Now let us compute the electric current helicity ${\lambda}$ which in the force-free dynamo case is given by
\begin{equation}
{\lambda}=\frac{\textbf{j}.\textbf{B}}{{\textbf{B}}^{2}}
\label{27}
\end{equation}
here by Maxwell equations the electric current $\textbf{j}$ is given by
\begin{equation}
\textbf{j}={\nabla}{\times}\textbf{B}
\label{28}
\end{equation}
From the closed two form $dB=0$ of the magnetic field yields
\begin{equation}
{B}_{x}:={\partial}_{y}A_{z}-{\partial}_{z}A_{y}
\label{29}
\end{equation}
\begin{equation}
{B}_{y}:=-{\partial}_{x}A_{z}+{\partial}_{z}A_{x}
\label{30}
\end{equation}
\begin{equation}
{B}_{z}:={\partial}_{x}A_{y}-{\partial}_{y}A_{x}
\label{31}
\end{equation}
From this definition one is able to determine the electric helicity and the magnetic helicity ${\cal{H}}$ \cite{3}
\begin{equation}
{\cal{H}}=\frac{1}{\sqrt{g}}[\textbf{A}.\textbf{B}]
\label{32}
\end{equation}
Since
\begin{equation}
j_{x}= {\partial}_{y}B_{z}
\label{33}
\end{equation}
and $B_{x}$ the electric helicity vanishes while the magnetic helicity can be expressed in terms of the magnetic vector potential as
\begin{equation}
{\partial}_{y}A_{x}=-B_{z}
\label{34}
\end{equation}
yields
\begin{equation}
A_{x}(y)=\int{B_{0}e^{-{\eta}{\lambda}^{2}t}y^{2}dy}
\label{35}
\end{equation}
which since the electric helicity ${\lambda}$ vanishes reduces to
\begin{equation}
A_{x}(y)=\int{B_{0}e^{-{\eta}{\lambda}^{2}t}y^{2}dy}=\frac{1}{3}B_{0}e^{-{\eta}{\lambda}^{2}t}y^{3}
\label{36}
\end{equation}
By considering that only $A_{z}$ and $B_{y}$ vanish and that the gauge vector magnetic potential is given by
\begin{equation}
{\nabla}.\textbf{A}=\frac{1}{{\eta}}{\phi}
\label{37}
\end{equation}
where ${\phi}(y,t)$ is the electric potential, the magnetic helicity may  be computed as
\begin{equation}
{\cal{H}}={B_{0}}^{2}y^{8}e^{-{\eta}{\lambda}^{2}t}
\label{38}
\end{equation}
whose electric potential is given by
\begin{equation}
{\phi}=5{\eta}y^{4}e^{-{\eta}{\lambda}^{2}t}
\label{39}
\end{equation}
This shows that the electric potential on non-compact Riemannian surfaces of negative curvature, which can be bound in the boundary of the Poincare discs, and decays in time. One also notes that in the ideal plasma case where the resistivity ${\eta}$ vanishes the gauge condition does not leads to the Weyl condition
\begin{equation}
{\nabla}.\textbf{A}=0
\label{40}
\end{equation}
unless at the center of the Poincare disc. The magnetic helicity also vanishes very fast as one approaches $y=0$. However this is forbidden in the Lobachevsky-Poincare plane, since there $y>0$. Thus not only electric potential but also magnetic helicity never vanish spatially at the Poincare disc, unless as $t\rightarrow{\infty}$.
\newpage
\section{Conclusions}
In general, fast dynamo are investigated in compact Riemannian manifolds, as has been shown by Arnold et al \cite{8} and by Chicone and Latushkin \cite{4}. In this paper, slow dynamos have been investigated in non-compact Riemannian manifolds. Here a toy model for a spatial hyperbolic section of a possible astrophysical dynamos in Lobachevsky plane is considered in 3D. This can serve as a disc dynamo in astrophysics or hyperbolic section of a cosmological model or even to investigate disc plasmas in laboratory as done by Fantoni et al. In the cosmological model the magnetic helicity can be investigated along with current helicity in the case of dynamos. These quantities are also useful in laboratory dynamos \cite{10}. Slow cosmic dynamos in plasmas can be obtained in laboratory as has been shown by Colgate et al \cite{11}. The investigation of restoring and viscous forces in the model may also serve as models for the geodynamos.
Note that here, despite of the fact that both magnetic and electric helicities vanish, the slow dynamo action in non-compact Riemannian manifolds of constant negative curvature. From the geodynamo and convection point of view in an interesting paper H Busse \cite{12} showed that the presence of curvilinear coordinates introduce new features on the rotating spherical shells that could be considered as Riemannian surface of positive Gaussian curvature. By considering that the plasma dynamo flow topology of the Poincare disc has a treshold in the growth rate of $|{\gamma}|\approx{0.022}$, performed in the laminar plasma dynamo experiment by Wang et al \cite{13}, from a $Rm\approx{210}$. From the expression ${\gamma}\approx{-{\eta}{\lambda}^{2}}$ one obtains that the electric current helicity can be determined as ${\lambda}\approx{2.34m^{-1}}$. In this computation, the inverse relation between the diffusion constant $\eta$ and the magnetic Reynolds number $Rm$ was used. All these physical applications make the model presented here useful in physical realistic situations and deserve further study.
\section{Acknowledgements}
I am very much indebt to J-Luc Thiffeault, Dmitry Sokoloff, Yu Latushkin and Rafael Ruggiero for reading for helpful discussions on the subject of this work. I appreciate financial  supports from UERJ and CNPq.

  \end{document}